\documentclass[sigconf]{acmart}
\def\BibTeX{{\rm B\kern-.05em{\sc i\kern-.025em b}\kern-.08emT\kern-.1667em\lower.7ex\hbox{E}\kern-.125emX}}
    
\usepackage{nicefrac}
\usepackage{siunitx}
\usepackage{array,framed}
\usepackage{booktabs}
\usepackage{
  color,
  float,
  epsfig,
  wrapfig,
  graphics,
  graphicx,
  subcaption
}
\usepackage{setspace}
\usepackage{latexsym,fancyhdr,url}
\usepackage{enumerate}
\usepackage{algorithm2e}
\usepackage{algpseudocode}
\usepackage{graphics}
\usepackage{xparse} 
\usepackage{xspace}
\usepackage{multirow}
\usepackage{csvsimple}
\usepackage{balance}
\usepackage{pifont}

\usepackage{
  tikz,
  pgfplots,
  pgfplotstable
}
\usepackage{hyperref}

\usetikzlibrary{
  shapes.geometric,
  arrows,
  external,
  pgfplots.groupplots,
  matrix
}

\pgfplotsset{compat=1.9}

\newcommand{\System}{\textsc{Pisco}\xspace}

\usepackage{mathtools,}
\usepackage[export]{adjustbox}
\usepackage{graphicx}
\usepackage{fancyhdr}
\usepackage[]{xcolor}

\DeclareMathAlphabet{\mathcal}{OMS}{cmsy}{m}{n}

\DeclareGraphicsExtensions{%
    .png,.PNG,%
    .pdf,.PDF,%
    .jpg,.mps,.jpeg,.jbig2,.jb2,.JPG,.JPEG,.JBIG2,.JB2}

\usepackage{xparse}
\newcommand{\bnm}{\begin{newmath}}
\newcommand{\enm}{\end{newmath}}

\newcommand{\bea}{\begin{eqnarray*}}%
\newcommand{\eea}{\end{eqnarray*}}%

\newcommand{\bne}{\begin{newequation}}
\newcommand{\ene}{\end{newequation}}

\newcommand{\bal}{\begin{newalign}}
\newcommand{\eal}{\end{newalign}}

\newenvironment{newalign}{\begin{align}%
\setlength{\abovedisplayskip}{4pt}%
\setlength{\belowdisplayskip}{4pt}%
\setlength{\abovedisplayshortskip}{6pt}%
\setlength{\belowdisplayshortskip}{6pt} }{\end{align}}

\newenvironment{newmath}{\begin{displaymath}%
\setlength{\abovedisplayskip}{4pt}%
\setlength{\belowdisplayskip}{4pt}%
\setlength{\abovedisplayshortskip}{6pt}%
\setlength{\belowdisplayshortskip}{6pt} }{\end{displaymath}}

\newenvironment{newequation}{\begin{equation}%
\setlength{\abovedisplayskip}{4pt}%
\setlength{\belowdisplayskip}{4pt}%
\setlength{\abovedisplayshortskip}{6pt}%
\setlength{\belowdisplayshortskip}{6pt} }{\end{equation}}

\newcounter{ctr}

\newcounter{mytable}
\def\mytable{\begin{centering}\refstepcounter{mytable}}
\def\endmytable{\end{centering}}

\newcounter{myfig}
\def\myfig{\begin{centering}\refstepcounter{myfig}}
\def\endmyfig{\end{centering}}

\newlength{\saveparindent}
\newlength{\saveparskip}
\newcommand{\E}{{\rm I\kern-.3em E}}

\renewcommand{\eqref}[1]{\mbox{Equation~(\ref{#1})}}

\def \part {part}

\renewcommand{\paragraph}[1]{\vspace*{6pt}\noindent\textbf{#1}\;}

\def \blackslug{\hbox{\hskip 1pt \vrule width 4pt height 8pt
    depth 1.5pt \hskip 1pt}}
\def \qed{\quad\blackslug\lower 8.5pt\null\par}

\newcounter{mynote}[section]

\newcommand\ignore[1]{}

\newcounter{rcnote}[section]

\newcounter{mrnote}[section]

\newcounter{fknote}[section]

\newcounter{anote}[section]

\DeclareMathSymbol{\mlq}{\mathord}{operators}{``}
\DeclareMathSymbol{\mrq}{\mathord}{operators}{`'}

\newcommand{\rhf}[2]{R_{f, \gamma}}

\DeclareDocumentCommand{\edist}{o o}{
  \ensuremath{
    \IfNoValueTF{#1}{{d}}{{\sf d}(#1,#2)}
  }
}

\newcommand{\olrk}[1]{\ifx\nursymbol#1\else\!\!\mskip4.5mu plus 0.5mu\left(\mskip0.5mu plus0.5mu #1\mskip1.5mu plus0.5mu \right)\fi}

\NewDocumentCommand{\indseq}{ O{1} O{r} }{{#1}\ldots {#2}}

\copyrightyear{2024}
\acmYear{2024}
\setcopyright{rightsretained}
\acmConference[CCS '24]{Proceedings of the 2024 ACM SIGSAC Conference on Computer and Communications Security}{October 14--18, 2024}{Salt Lake City, UT, USA}
\acmDOI{10.1145/3658644.3691381}
\acmISBN{979-8-4007-0636-3/24/10}

\begin{document}
\fancyhead{}
\def\thetitle{Different Victims, Same Layout: Email Visual Similarity Detection for Enhanced
Email Protection}
\title{\thetitle}

\author{Sachin Shukla}
\affiliation{Cisco Talos, CA, USA}

\author{Omid Mirzaei}
\affiliation{Cisco Talos, VA, USA}

\date{}
\begin{abstract}

In the pursuit of an effective spam detection system, the focus has often been on identifying known spam patterns either through rule-based detection systems or machine learning (ML) solutions that rely on keywords. However, both systems are susceptible to evasion techniques and zero-day attacks that can be achieved at low cost. Therefore, an email that bypassed the defense system once can do it again in the following days, even though rules are updated or the ML models are retrained. The recurrence of failures to detect emails that exhibit layout similarities to previously undetected spam is concerning for customers and can erode their trust in a company. Our observations show that threat actors reuse email kits extensively and can bypass detection with little effort, for example, by making changes to the content of emails. In this work, we propose an email visual similarity detection approach, named \System, to improve the detection capabilities of an email threat defense system. We apply our proof of concept to some real-world samples received from different sources. Our results show that email kits are being reused extensively and visually similar emails are sent to our customers at various time intervals. Therefore, this method could be very helpful in situations where detection engines that rely on textual features and keywords are bypassed, an occurrence our observations show happens frequently.

\end{abstract}

\keywords{Phishing, Spam Detection, Email Kit, Visual Similarity, Image Embedding}
\maketitle
\section{Introduction}
\label{sec:intro}
In the digital age, email stands as a cornerstone of communication, facilitating a myriad of personal and professional interactions. However, as the volume of email traffic continues to soar, so does the proliferation of spam messages. These unsolicited emails inundate inboxes, erode productivity, and pose significant security risks. Email is still the primary attack vector being abused by threat actors for initial access to their targets \cite{Talos_Report_2023}. The battle against email spam has witnessed substantial progress through the deployment of sophisticated detection systems.

As security measures improve and detection mechanisms become more sophisticated, cybercriminals are quick to adapt, devising elusive methods to bypass these defenses. Specifically, with the inexpensive phishing-as-a-service kits (e.g., Caffeine \cite{Mccabe22}, EvilProxy \cite{Resecurity22}, and NakedPages \cite{CloudSek22}) that are available, bad actors can get access to a wide range of capabilities. Most phishing kits are found to be distributed and reused in whole or in part \cite{Feller21, MSFINT21}. On the other hand, email kits offer attackers an easy way to curate and send emails to a larger audience, maximizing the impact of their email campaigns\footnote{Sha256 of a recently discovered email kit used by a North Korean threat actor: bb9c0396a61fa16d8c482a4a17e520fae908aa826e54243da6473494fa5f2305}. These kits often provide a number of default email templates to make the process even smoother and faster. 

Recently, defense strategies that rely on visual similarity \cite{abdelnabi2020visualphishnet,liu2022inferring,lin2021phishpedia} have become some of the most powerful and successful methods for detecting phishing pages and countering phishing attacks. These mechanisms necessitate the creation of a reference collection featuring visual elements, such as screenshots and logos, of widely recognized brands obtained from legitimate web pages. They are adept at identifying potential phishing attacks by spotting deceitful sites that bear a strong visual similarity to established brands (e.g., PayPal), yet operate under disparate domain names. However, these methods have not been applied to emails directly in the literature. 

\begin{figure*}[t!]
    \subfloat[Timestamp: 10/02/2023]{%
        \includegraphics[width=.42\textwidth, fbox]{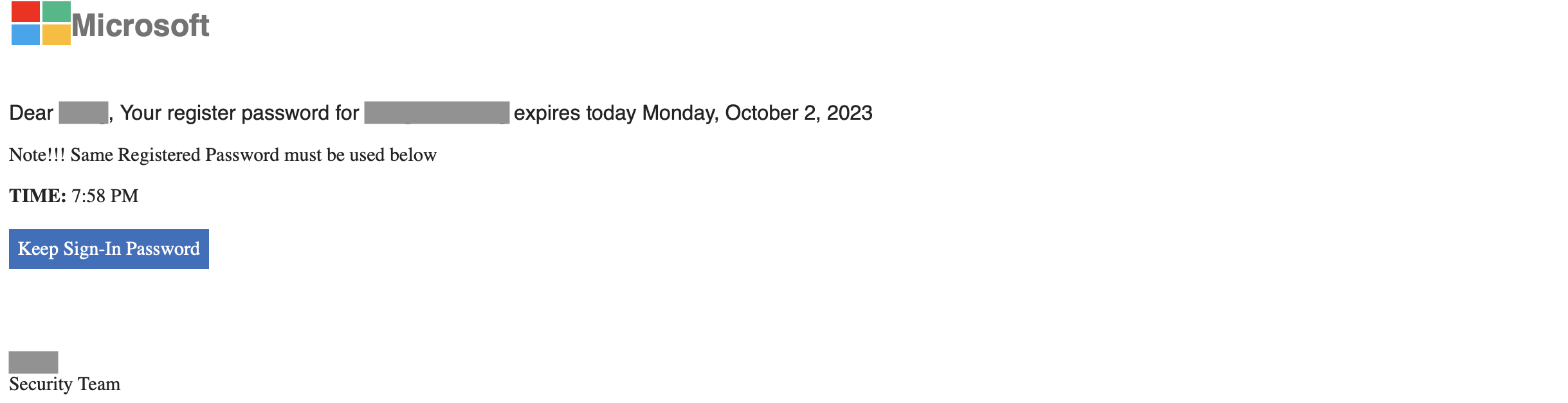}%
    }\hfill
    \subfloat[Timestamp: 10/05/2023]{%
        \includegraphics[width=.42\textwidth, fbox]{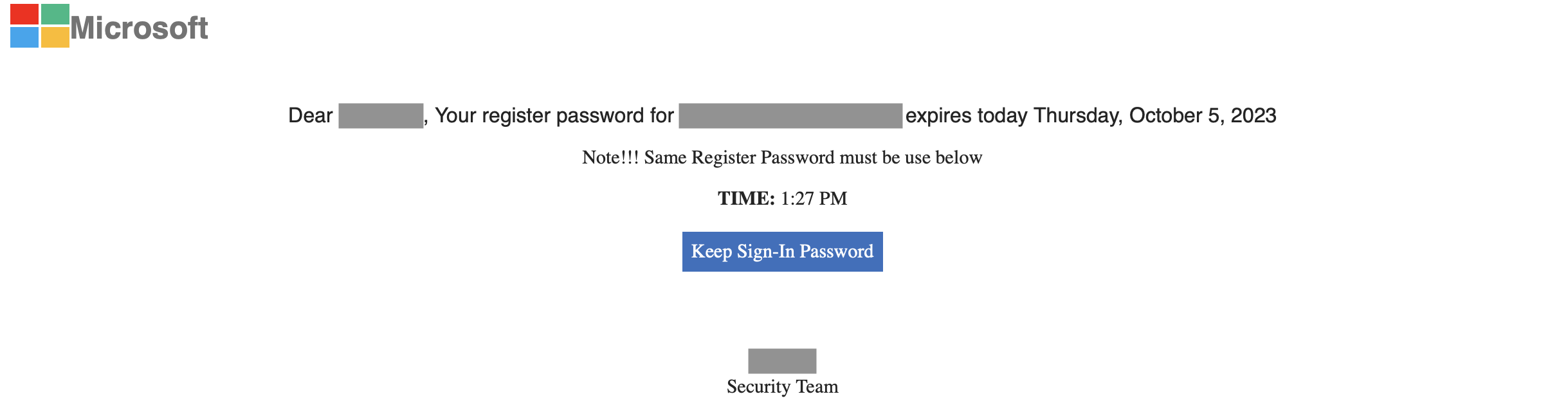}%
    }
    \\
    \subfloat[Timestamp: 10/20/2023]{%
        \includegraphics[width=.42\textwidth, fbox]{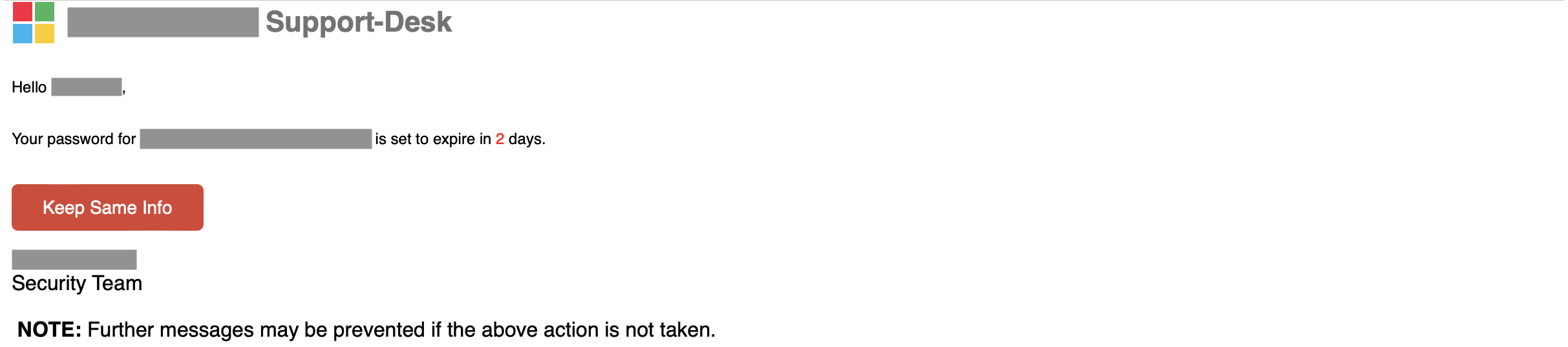}%
    }\hfill
    \subfloat[Timestamp: 10/24/2023]{%
        \includegraphics[width=.42\textwidth, fbox]{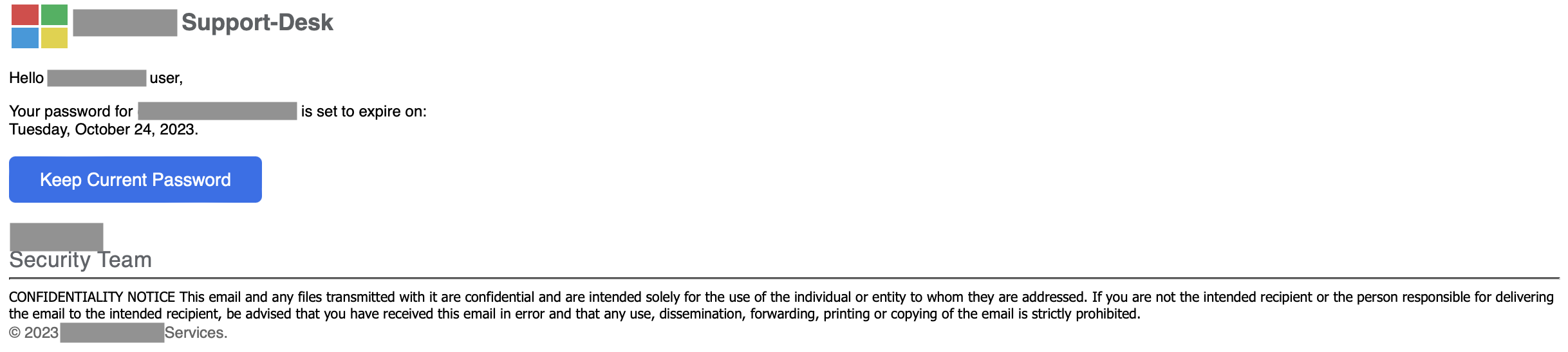}%
    }
    \caption{Four visually similar phishing emails sent to our distinct customers at different time intervals.}
    \label{fig:cluster_example}
\end{figure*}

Our observations show that email kits are reused extensively in the wild (see Figure \ref{fig:cluster_example} as an example), and several emails with layout similarities are observed in our telemetry at different time intervals. We believe this can happen for two main reasons: 1) An old threat actor changing their infrastructure but reusing the same email kit as a few days ago, or 2) Different threat actors using the same email kit. Email kits provide different features that enable threat actors to add additional elements (e.g., logos) to emails. They also allow them to re-paraphrase specific parts of emails to evade detection systems that rely on keywords. Therefore, email kit reuse introduces significant challenges to conventional defense mechanisms such as rule-based detection engines and even ML-based solutions that rely on textual features and are not robust against zero-day attacks.

In this work, we propose an email visual similarity detection approach, named \System, to enable the email defense system convict new emails that resemble to any historic spam emails. On the backend, our approach renders the HTML source of each email we receive from different sources in a client, and next, takes a screenshot of the fully rendered email. It then embeds the screenshot’s image to a numerical representation via deep learning, and clusters the embedding vectors into different groups to create a knowledge base. On the frontend, it embeds each new email that is going to be delivered to the inboxes in the same way and returns a verdict for the email based on the assigned cluster.

\section{approach}
\label{sec:approach}

This section presents our approach to detect visually similar emails via image embedding. The system begins with the collection of email screenshots, followed by a series of image processing steps to enhance the quality and relevance of the captured images. The processed images are then embedded into vector representations using state-of-the-art pre-trained models and are stored in a database. Finally, an image similarity detection and retrieval mechanism is employed to identify visually similar emails within the database to any given email.

\subsection{Image Capturing}
\label{sec:image_capturing}

The initial step in our pipeline involves capturing email screenshots. This process is crucial for extracting visual information that represents the appearance and layout of the email. The system utilizes automated tools to render the HTML source of each email in an email client, and to capture relevant portions of the fully rendered email.

Our approach parses each email in the first step to retrieve its contents in accordance with the RFC822 standard \cite{RFC822}. Then, email banners are removed, and the email is re-written. Banners are added to emails by different services, and they serve different purposes. For example, security warnings are typically added in a banner to the message by email security gateways. Once the banner is removed, the message is rendered in the Thunderbird email client via an X virtual framebuffer, known as Xvfb. When fully loaded, the message screenshot is taken and stored for image processing.

\subsection{Image Processing}
\label{sec:image_processing}

To standardize and enhance the quality of captured images, a set of image processing procedures is applied. These procedures include normalization techniques to adjust lighting and color variations, and cropping to focus on the relevant content. By incorporating these processing steps, the system aims to create a consistent and normalized data for improved image embedding. In particular, we leveraged the following image processing techniques:

\subsubsection{Normalization}
\label{sec:normalization}

In image processing, normalization typically involves modifying pixel values to adhere to a particular range or distribution, often with the goal of enhancing contrast or standardizing intensity values. Common normalization techniques, such as histogram equalization and contrast stretching, reorganize pixel intensities across a wider spectrum, resulting in a visually appealing image. We rely on the NORM\_MINMAX method of the OpenCV library to normalize the emails' screenshots \cite{OpenCV_normalization}.

\subsubsection{Sharpening}
\label{sec:normalization}

The primary objective of image sharpening is to augment the high-frequency elements within an image, thereby accentuating edges and intricate details. This enhancement is frequently accomplished through the application of filters, such as the unsharp mask or sharpening filter, which operate by highlighting variations in intensity among neighboring pixels, ultimately intensifying the image's edges. In our research, we have adopted an adaptive approach to image sharpening \cite{Wand_sharpen}. In contrast to conventional methods that uniformly enhance the sharpness of an entire image, adaptive sharpening tailors the degree of sharpness according to the content and features found in distinct areas.

\subsubsection{Cropping}
\label{sec:cropping}

In the context of image processing, cropping involves eliminating outer segments of an image to preserve only the intended section. This technique includes choosing a designated area of interest and discarding the remaining portions. In this work we leverage image cropping to discard the email's header and to remove the redundant white spaces surrounding the email body.

\subsection{Image Embedding}
\label{sec:image_embedding}

Image embedding serves as a crucial step in our pipeline, transforming processed images into numerical vector representations. By utilizing pre-trained models, the system extracts meaningful features and encodes them into a high-dimensional vector space. In our application context, this embedding vector captures the numerical representation of the visual elements embedded within each email. If two emails yield similar image embedding vectors, it suggests that their screenshots exhibit comparable visual characteristics. This similarity serves as an indicator that the emails have analogous visual elements and share common templates or visual layouts. Therefore, the embedding process streamlines the effective comparison and retrieval of visually similar emails. In our work, we have leveraged a pre-trained CLIP model, developed by OpenAI \cite{radford2021learning}, to embed the screenshot of emails.

\subsection{Image Retrieval and Similarity Detection}
\label{sec:image_similarity_detection}

Once a rich database of email embeddings is created from different sources, the incoming emails are compared with historic ones and are convicted if a high visual similarity exists with spam messages. The image similarity detection and retrieval system should be both fast and reliable to be applicable to this application context. In our work, we rely on a fast open-source vector database, called Milvus \cite{Milvus}, that relies on nearest neighbor search algorithm and has been widely used in industry for similar purposes. Nearest neighbor search algorithms can be used to index a very large number of email embedding vectors. If the ultimate goal is to group visually similar emails together, you can achieve this by leveraging either a clustering algorithm or a similarity threshold.

\section{Preliminary Results}
\label{sec:results}

\begin{figure}[t]
\centering\includegraphics[width=\columnwidth]{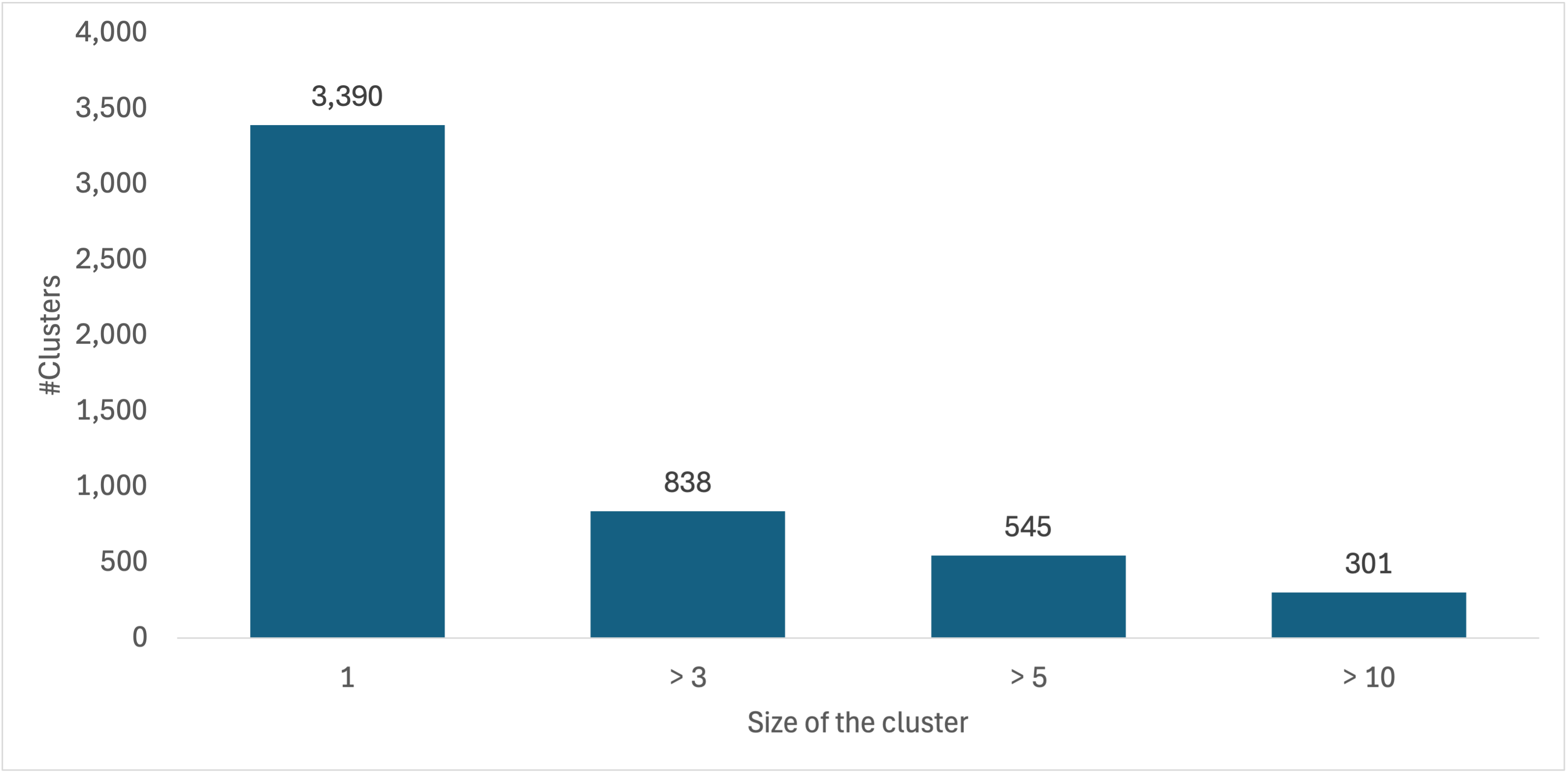}
\caption{The clustering result on a subset of corpus emails.}
\label{fig:clustering_result}
\end{figure}

This section summarizes our key findings. We ran our prototype on emails we received from different sources in our corpus within a span of one month (April 2024). We used some pre-filtering criteria to down-sample messages and some other criteria to discard messages that did not have any visual elements. This yielded around 116K emails in the studied time window. From the number of messages we processed, we detected 20,215 clusters of visually similar emails. Figure \ref{fig:clustering_result} shows that only 3,390 clusters (2.92\%) had one email, and the remaining ones had at least two emails that share some visual similarity. We manually inspected all clusters and confirmed that no false positives existed from a similarity detection perspective. 

\begin{figure}[t]
\centering\includegraphics[width=\columnwidth]{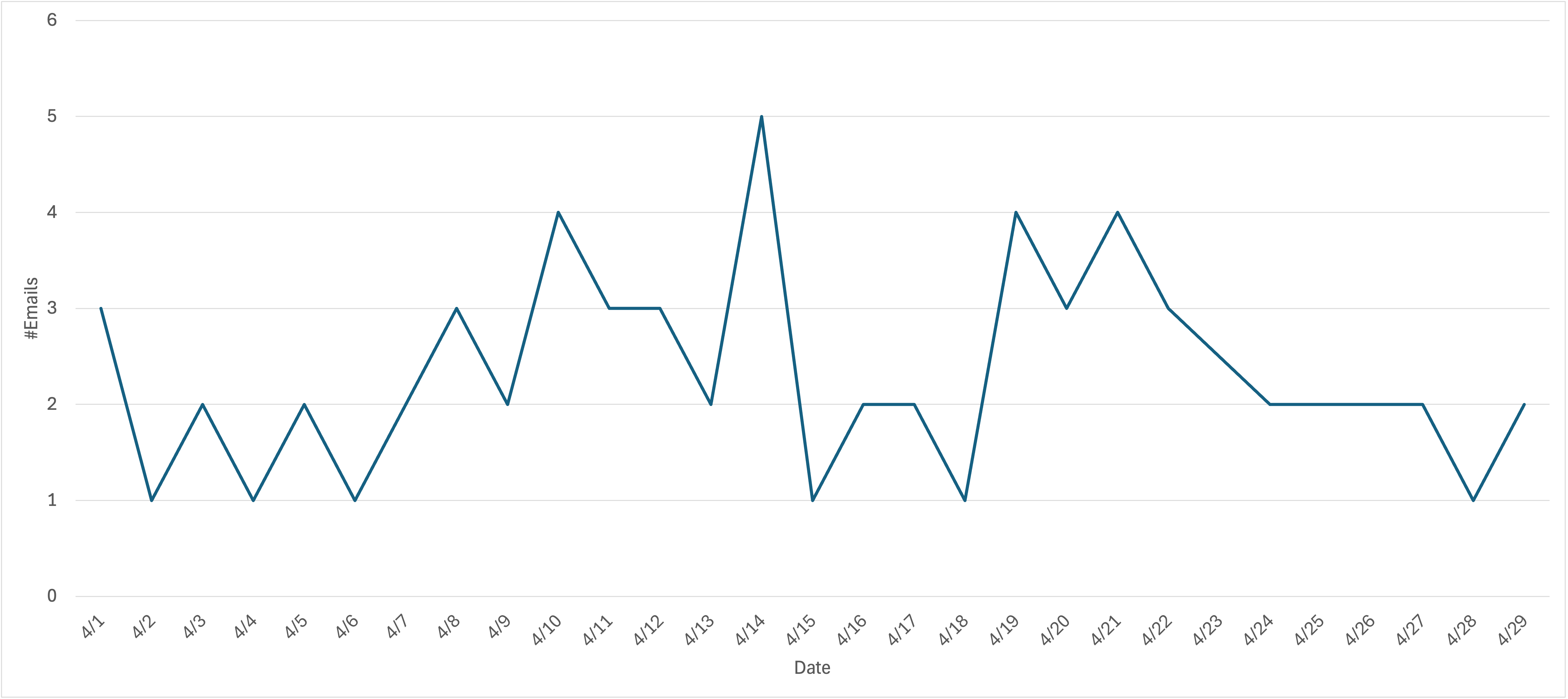}
\caption{The cluster lifespan of an adult spam campaign, showcasing visually similar emails submitted to our corpus at various times throughout April 2024.}
\label{fig:adultspam_cluster_lifespan}
\end{figure}

Experimental results show that most clusters have medium to long life spans (see Figure \ref{fig:adultspam_cluster_lifespan} as an example), i.e., emails in such clusters were received at different time intervals across weeks or sometimes even months. This confirms that email kits are shared extensively and are reused at different time intervals. Therefore, if historic emails with visual elements are embedded and stored in a database, they can be used at a later time for enhanced protection.

\section{Conclusion and Future Work}
\label{sec:conclusion}

Our prototype analysis of emails received over one month revealed that a substantial portion of our corpus emails were visually similar (i.e, they had similar layouts), forming 20,215 clusters. The majority of these clusters consisted of multiple emails, indicating extensive sharing and reuse of email kits over time. Our findings suggest that incorporating historic emails with visual elements into a database could significantly improve future protection efforts.

Although the immediate use case of our system is to enhance the protection capabilities of our email threat defense product, we have several ongoing projects that apply this approach to other use cases, including visual search for improved threat intelligence, automatic message labeling, and email campaign tracking.

\section{Acknowledgement}
\label{sec:acknowledgement}

A patent application for this work has been filed with the United States Patent and Trademark Office (USPTO).
\bibliographystyle{ACM-Reference-Format}
\bibliography{references}


\end{document}